\documentclass[a4paper]{jpconf}
\usepackage{graphicx}
\usepackage{amsmath}
\usepackage{amsfonts}
\usepackage{amssymb}
\usepackage{graphicx}

\begin{document}
\title{Vacuum properties of open charmed mesons in a chiral symmetric model}

\author{Walaa I. Eshraim}

\address{Institute for Theoretical Physics, Goethe-University, Max-von-Laue-Str. 1,
\\60438 Frankfurt am Main, Germany}

\ead{weshraim@th.physik.uni-frankfurt.de}

\begin{abstract}
We present a $U(4)_R \times U(4)_L$ chirally symmetric model, which in addition to scalar and pseudoscalar mesons also includes vector and axial-vector mesons.
A part from the three new parameters pertaining to the charm degree of freedom, the parameters of the model are fixed from the $N_f=3$ flavor sector. We calculate
 open charmed meson masses and the weak decay constants of nonstrange open charm $D$ and strange open charm $D_S$. We also evaluate the (OZI-dominant) strong decays of open charmed mesons. The results are turn out to be in quantitative agreement with experimental data.
\end{abstract}

\section{Introduction}

Open charmed mesons, composite states of charm quark (c) and up
(u), down (d), or strange (s) antiquark, were observed two years
later than the discovery of the $J/\psi$ particle in 1974. Since
that time, the study of charmed meson spectroscopy and decays has
made significant experimental \cite{BESIII, experiments, D mesons}
and theoretical process \cite{models1, models2, lattice, hqet}. We
show in the present work that how the original $SU(3)$ flavor
symmetry of hadrons can be extended to $SU(4)$ in the framework of
a chirally symmetric model with charm as an extra quantum number.
Note that, chiral symmetry is strongly explicitly broken by the
current charm quark mass.

The development of an effective hadronic Lagrangian plays an important role in the description of the masses and the interactions of low-lying hadron
resonances \cite{Amsler}. To this end, we developed the so-called extended Linear Sigma Model (eLSM) in which (pseudo)scalar and (axial-)vector
 $q\overline{q}$ mesons and additional scalar and pseudoscalar glueball fields are the basic degrees of freedom. The eLSM has already shown success in
 describing the vacuum phenomenology of the nonstrange-strange mesons \cite{Lenaghan,Bando,2D,3D,3Dp}. The eLSM emulates the global symmetries of the QCD Lagrangian; the
 global chiral symmetry (which is exact in the chiral limit), the discrete C, P, and T symmetries, and the classical dilatation (scale) symmetry. When
 working with colorless hadronic degrees of freedom, the local color symmetry of QCD is automatically preserved. In eLSM the global chiral symmetry is
 explicitly broken by non-vanishing quark masses and quantum effects \cite{Hooft}, and spontaneously by a non-vanishing expectation value of the quark
 condensate in the QCD vacuum \cite{Vafa}. The dilatation symmetry is broken explicitly by the logarithmic
term of the dilaton potential, by the mass terms, and by the
$U(1)_A$ anomaly.

In these proceedings, we present the outline of the extension of
the eLSM from the three-flavor case to the four-flavor case
including the charm quark \cite{WFDP, WPr1, WPr2}. Most parameters
of our eLSM are taken directly from Ref.\cite{3D} where the
nonstrange-strange mesons were considered. There are only new
three parameters pertaining to the charm degree of freedom. We
compute open charmed meson masses, the weak decay constants of the
pseudoscalar $D$ and $D_s$ mesons, and
 the (OZI-dominant) strong decays
of open charmed mesons.

\section{The $U(4)_R \times U(4)_L$ Linear Sigma Model}

In Refs.\cite{WFDP, WPr1, WPr2}, we presented the outline of the
extension of the eLSM from the three-flavor case to the
four-flavor case including charm quark. In this extension, we
introduced the (pseudo)scalar
and (axial-)vector meson fields in terms of $%
4\times 4$ (instead of $3\times 3$) matrices, which the charmed
mesons appear in the fourth row and fourth column as follows: The
matrix of pseudoscalar fields $P$ (with quantum numbers
$J^{PC}=0^{-+}$) reads
\begin{equation}
P=\frac{1}{\sqrt{2}}\left(
\begin{array}
[c]{cccc}%
\frac{1}{\sqrt{2}}(\eta_{N}+\pi^{0}) & \pi^{+} & K^{+} & D^{0}\\
\pi^{-} & \frac{1}{\sqrt{2}}(\eta_{N}-\pi^{0}) & K^{0} & D^{-}\\
K^{-} & \overline{K}^{0} & \eta_{S} & D_{S}^{-}\\
\overline{D}^{0} & D^{+} & D_{S}^{+} & \eta_{c}%
\end{array}
\right)  \text{ ,} \label{p}%
\end{equation}
and the matrix of scalar fields $S$ (with quantum numbers $J^{PC}=0^{++}$) reads%
\begin{equation}
S=\frac{1}{\sqrt{2}}\left(
\begin{array}
[c]{cccc}%
\frac{1}{\sqrt{2}}(\sigma_{N}+a_{0}^{0}) & a_{0}^{+} &
K_{0}^{\ast+} &
D_{0}^{\ast0}\\
a_{0}^{-} & \frac{1}{\sqrt{2}}(\sigma_{N}-a_{0}^{0}) &
K_{0}^{\ast0} &
D_{0}^{\ast-}\\
K_{0}^{\ast-} & \overline{K}_{0}^{\ast0} & \sigma_{S} & D_{S0}^{\ast-}\\
\overline{D}_{0}^{\ast0} & D_{0}^{\ast+} & D_{S0}^{\ast+} & \chi_{c0}%
\end{array}
\right)  \text{ ,}%
\end{equation}
 which are used to construct the matrix $\Phi=S+iP$. In the
 pseudoscalar sector there are: an open charmed state $D^{0,\pm}$,
 open strange-charmed states $D^\pm_S$, and a hidden charmed ground
 state $\eta_C(1S)$. In the scalar sector there are open charmed
 $D^{\ast\,0,\pm}_0$ and strange charmed meson $D^{\ast\,\pm}_{S0}$ which are
 assigned to $D^\ast_0(2400)^{0,\pm}$ and $D_{S0}^\ast(2317)^\pm$,
 respectively.\\
We now turn to the vector sector. The matrix $V^{\mu}$ which
includes the
vector degrees of freedom is:%
\begin{equation}
V^{\mu}=\frac{1}{\sqrt{2}}\left(
\begin{array}
[c]{cccc}%
\frac{1}{\sqrt{2}}(\omega_{N}+\rho^{0}) & \rho^{+} &
K^{\ast}(892)^{+} &
D^{\ast0}\\
\rho^{-} & \frac{1}{\sqrt{2}}(\omega_{N}-\rho^{0}) &
K^{\ast}(892)^{0} &
D^{\ast-}\\
K^{\ast}(892)^{-} & \bar{K}^{\ast}(892)^{0} & \omega_{S} & D_{S}^{\ast-}\\
\overline{D}^{\ast0} & D^{\ast+} & D_{S}^{\ast+} & J/\psi
\end{array}
\right)  ^{\mu}\text{ ,}%
\end{equation}
where the nonstrange-charmed fields $D^{\ast 0},\,D^{\ast\pm }$
correspond to $\overline{q}q$ resonances $D^{\ast }(2007)^{0}$ and
$D^{\ast }(2010)^{\pm }$, respectively, while the strange-charmed
$D_{0}^{\ast\pm}$ is assigned to the resonance $D_{0}^{\ast\pm}$
(with mass $m_{D_{0}^{\ast\pm}}$), and there is the lowest vector
charmonium state $J/\psi(1S)$. \\
The matrix $A^{\mu}$ describing the axial-vector degrees of
freedom is given
by:%
\begin{equation}
A^{\mu}=\frac{1}{\sqrt{2}}\left(
\begin{array}
[c]{cccc}%
\frac{1}{\sqrt{2}}(f_{1,N}+a_{1}^{0}) & a_{1}^{+} & K_{1}^{+} & D_{1}^{0}\\
a_{1}^{-} & \frac{1}{\sqrt{2}}(f_{1,N}-a_{1}^{0}) & K_{1}^{0} & D_{1}^{-}\\
K_{1}^{-} & \bar{K}_{1}^{0} & f_{1,S} & D_{S1}^{-}\\
\bar{D}_{1}^{0} & D_{1}^{+} & D_{S1}^{+} & \chi_{c,1}%
\end{array}
\right)  ^{\mu}\,,
\end{equation}
where the open charmed mesons $D_{1}$ and $D_{S1}$ are assigned to
$D_{1}(2420)$ and $D_{S1}(2536)$, respectively, whereas the
charm-anticharm state $\chi _{c1}$ corresponds to the
$c\overline{c}$ resonance $\chi _{c1}(1P)$. From the matrices
$V^{\mu}$ and $A^{\mu}$ we construct the left-handed and
right-handed vector fields $L^{\mu}=V^{\mu}+A^{\mu}$ and
$R^{\mu}=V^{\mu }-A^{\mu}$, respectively. \\
\indent The explicit form of the eLSM
Lagrangian for $N_{f}=4$  is analogous to the case
$N_{f}=3$ of Ref. \cite{3D,3Dp} (but has an additional term $-2\,\mathrm{Tr}%
[E\Phi ^{\dagger }\Phi]$):
\begin{align}
\mathcal{L}  & =\frac{1}{2}(\partial _{\mu
}G)^{2}-V_{dil}(G)+\mathrm{Tr}[(D^{\mu}\Phi)^{\dagger}(D^{\mu
}\Phi)]-m_{0}^{2}\left(  \frac{G}{G_{0}}\right)
^{2}\mathrm{Tr}(\Phi
^{\dagger}\Phi)-\lambda_{1}[\mathrm{Tr}(\Phi^{\dagger}\Phi)]^{2}\nonumber\\
& -\lambda
_{2}\mathrm{Tr}(\Phi^{\dagger}\Phi)^{2}+\mathrm{Tr}[H(\Phi+\Phi^{\dagger
})]
 +\mathrm{Tr}\left\{  \left[  \left(  \frac{G}{G_{0}}\right)  ^{2}%
\frac{m_{1}^{2}}{2}+\Delta\right]  \left[
(L^{\mu})^{2}+(R^{\mu})^{2}\right]
\right\} \nonumber\\
&  -\frac{1}{4}\mathrm{Tr}[(L^{\mu\nu})^{2}+(R^{\mu\nu})^{2}%
]-2\,\mathrm{Tr}[E\,\Phi^{\dagger}\Phi]
+c(\mathrm{det}\Phi
-\mathrm{det}\Phi^{\dagger})^{2}+ic_{\tilde{G}\Phi}\tilde{G}(\mathrm{det}\Phi
-\mathrm{det}\Phi^{\dagger})\nonumber\\
&  +i\frac{g_{2}}{2}\{\mathrm{Tr}(L_{\mu\nu}[L^{\mu},L^{\nu}])+\mathrm{Tr}%
(R_{\mu\nu}[R^{\mu},R^{\nu}])\}+\frac{h_{1}}{2}\mathrm{Tr}(\Phi^{\dagger}%
\Phi)\mathrm{Tr}[(L^{\mu})^{2}+(R^{\mu})^{2}]\nonumber\\
&+h_{2}\mathrm{Tr}[(\Phi R^{\mu })^{2}+(L^{\mu}\Phi)^{2}]
+2h_{3}\mathrm{Tr}(\Phi R_{\mu}\Phi^{\dagger}L^{\mu})+ \ldots\,,\,
\label{lag}%
\end{align}

where the field $G$ denotes the dilaton field and its potential \cite{Salomone}
reads
\begin{equation}
V_{dil}(G)=\frac{1}{4}\frac{m_{G}^2}{\Lambda_G^2}\bigg[G^4\rm{ln}\bigg(\frac{G}{\Lambda_G}\bigg)-\frac{G^4}{4}\bigg],
\end{equation}
in which the parameter $\Lambda_{G}\sim N_C\,\Lambda_{QCD}$ sets the
energy scale of the gauge theory. The dilaton potential breaks the dilatation symmetry explicitly. $D^\mu\Phi\equiv\partial^\mu\Phi-ig_1 (L^\mu \Phi-\Phi
R^\mu)$ is the covariant derivative;
$L^{\mu\nu}\equiv \partial^\mu L^\nu -
\partial^\nu L^\mu$, and $R^{\mu\nu}\equiv\partial^\mu R^\nu -
\partial^\nu R^\mu$ are the left-handed and
right-handed field strength tensors.
 In eLSM Lagrangian (\ref{lag}) the dots
refer to further chirally invariant terms listed in Ref.\
\cite{3D}: these terms do not affect the masses and decay widths
studied in the present work and we therefore omitted them. The
term $ic_{\tilde{G}\Phi}\tilde{G}\left(
{\textrm{det}}\Phi-{\textrm{det}}\Phi^{\dag}\right)$ describes the
interaction between the pseudoscalar glueball
$\widetilde{G}\equiv|gg>$ and (pseudo-)scalar mesons, which is
used to study the phenomenology of the pseudoscalar glueball in
the case of $N_f=3$ \cite{Eshraim}.
 The terms $\mathrm{Tr}[H(\Phi +\Phi
^{\dagger })]$ with $H=1/2\,\text{diag}\{h_{0N},\,h_{0N},\,\sqrt{2}h_{0S},\,\sqrt{2}%
h_{0C}\}$, $-2\,\mathrm{Tr}%
[E\Phi ^{\dagger }\Phi]$ with $E=\text{diag}\{\varepsilon
_{N},\,\varepsilon _{N},\,\varepsilon _{S},\,\varepsilon
_{C}\},\,\varepsilon _{i}\propto m_{i}^{2},\,\varepsilon
_{N}=\varepsilon _{S}=0$, and $\mathrm{Tr}\left[ \Delta (L^{\mu
}{}^{2}+R^{\mu }{}^{2})\right]$ with $\delta =\text{diag}\{\delta
_{N},\,\delta _{N},\,\delta _{S},\,\delta _{C}\},\,\delta _{i}\sim
m_{i}^{2},\,\delta _{N}=\delta _{S}=0$, break chiral symmetry due
to nonzero quark masses and are especially important for mesons
containing the charm quark. When $m_{0}^{2}<0$ spontaneous
symmetry breaking occurs and the scalar-isoscalar fields condense
as well as the glueball field $G=G_0$. To implement this breaking
we shift $\sigma _{N}$, $\sigma _{S}$, $G$, and $\chi _{C0}$ by
their respective vacuum expectation values
$\phi_N,\,\phi_S,\,G_0,$ and $\phi_C$ \cite{WFDP, WPr1, WPr2} as
\begin{align}
\sigma_N \rightarrow \sigma_N + \phi_N,\,\,
\sigma_S \rightarrow \sigma_S+\phi_S\;\,,
G \rightarrow G+G_0,\,\, \rm{and}\,\, \chi_{C0} \rightarrow \chi_{C0}+\phi_C\,.
\end{align}

Most of the parameters of the model were already fixed in the
three-flavor study of Ref. \cite{3D}. Only three new parameters
appear and all of them are related to the bare mass of the charm
quark. They were determined in Ref. \cite{WFDP} through a fit to
the masses of charmed mesons. As an outcome, the charm-anticharm
condensate is sizable, $\phi_C =178\pm 28$  MeV.

\section{Results}
The weak-decay constants of the pseudoscalar open charmed mesons
$D$ and $D_{S}$ \cite{WFDP, WPr1, WPr2} are
$$f_{D}=\frac{\phi_{N}+\sqrt{2}\phi_{C}}{\sqrt{2}Z_{D}}=(254\pm17)\text{
MeV },\,\,\,\,\,\,\,\,\,\,\,
f_{D_{S}}=\frac{\phi_{S}+\phi_{C}}{Z_{D_{S}}}=(261\pm17)\text{ MeV
}\,,$$ where the experimental values \cite{PDG} are
$$f_{D}=(206.7\pm8.9)\,\text{MeV},\,\,\,\,\,\,\,\,\,\,\,\,\,\,\,\,\,\,\,\,\,\,\,\,f_{D_{s}}=(260.5\pm5.4)\rm{MeV}\,.$$

The results for the open charmed meson masses are reported in Table \textbf{1} \cite{WFDP, WPr1}. They have been obtained through a fit to experimental data\\
\begin{center}
\textbf{Table\thinspace1}\thinspace:\thinspace\ Masses of open charmed meson.%

\begin{tabular}
[c]{|c|c|c|c|c|}\hline
Resonance &  Our Value [MeV] & Experimental Value[MeV] \\\hline
$D^{0}$ & $1981\pm73$ & $1864.86\pm0.13$ \\\hline
$D_{S}^{\pm}$  & $2004\pm74$ & $1968.50\pm0.32$\\\hline
$D_{0}^{\ast}(2400)^{0}$  & $2414\pm77$ & $2318\pm29$ \\\hline
$D_{S0}^{\ast}(2317)^{\pm}$ & $2467\pm76$ & $2317.8\pm0.6$ \\\hline
$D^{\ast}(2007)^{0}$  & $2168\pm70$ & $2006.99\pm0.15$\\\hline
$D_{s}^{\ast}$ & $2203\pm69$ & $2112.3\pm0.5$\\\hline
$D_{1}(2420)^{0}$ & $2429\pm63$ & $2421.4\pm0.6$ \\\hline
$D_{S1}(2536)^{\pm}$ & $2480\pm63$ &
$2535.12\pm0.13$ \\\hline
\end{tabular}
\end{center}

The results of (OZI-dominant) strong decay widths of the open
charmed mesons described by the resonances $D_{0}^*,$ $D^{\ast},$
and $D_{1}$ are summarized in Table \textbf{2} \cite{WFDP, WPr2}.

\begin{center}
\textbf{Table\thinspace2:}\thinspace\thinspace\ Decay widths of charmed
mesons
%\bigskip%
\begin{tabular}
[c]{|c|c|c|}\hline
Decay Channel & Theoretical result [MeV] & Experimental result [MeV]\\\hline
$D_{0}^{\ast}(2400)^{0}\rightarrow D\pi$ &
$139_{-114}^{+243}$ & full width $\Gamma=267\pm40$\\\hline
$D_{0}^{\ast}(2400)^{+}\rightarrow D\pi$ &
$51_{-51}^{+182}$ &  full width: $\Gamma=283\pm24\pm34$\\\hline
$D^{\ast}(2007)^{0}\rightarrow D^{0}\pi^{0}$ & $0.025\pm0.003$ & $<1.3$\\\hline
$D^{\ast}(2007)^{0}\rightarrow D^{+}\pi^{-}$ & $0$ & not seen\\\hline
$D^{\ast}(2010)^{+}\rightarrow D^{+}\pi^{0}$ & $0.018_{-0.003}^{+0.002}$ &
$0.029\pm0.008$\\\hline
$D^{\ast}(2010)^{+}\rightarrow D^{0}\pi^{+}$ & $0.038_{-0.004}^{+0.005}$ & $0.065\pm0.017$\\\hline
$D_{1}(2420)^{0}\rightarrow D^{\ast}\pi$ &
$65_{-37}^{+51}$ &  full width: $\Gamma=27.4\pm
2.5$\\\hline
$D_{1}(2420)^{0}\rightarrow D^{0}\pi\pi$ & $0.59\pm0.02$ & seen\\\hline
$D_{1}(2420)^{0}\rightarrow D^{+}\pi^{-}\pi^{0}$ & $0.21_{-0.015}^{+0.01}$ &
seen\\\hline
$D_{1}(2420)^{0}\rightarrow D^{+}\pi^{-}$ & $0$ & not seen; $\Gamma(D^{+}%
\pi^{-})/\Gamma(D^{\ast+}\pi^{-})<0.24$\\\hline
$D_{1}(2420)^{+}\rightarrow D^{\ast}\pi$ &
$65_{-36}^{+51}$ & full width: $\Gamma=25\pm
6$\\\hline
$D_{1}(2420)^{+}\rightarrow D^{+}\pi\pi$ & $0.56\pm0.02$ & seen\\\hline
$D_{1}(2420)^{+}\rightarrow D^{0}\pi^{0}\pi^{+}$ & $0.22\pm0.01$ &
seen\\\hline
$D_{1}(2420)^{+}\rightarrow D^{0}\pi^{+}$ & $0$ & not seen\\\hline
\end{tabular}

\end{center}

\section{Conclusion}

In this work we have presented the outline of the extension of the eLSM from the three-flavor case to the four-flavor case including the charm quark
has been presented. Most parameters are determined in the low-energy study for the nonstrange-strange sector \cite{3D}. Three new unknown
parameters have been fixed in a fit to the experimental values (details are presented in Ref.\cite{WFDP}). The weak decay constants of nonstrange charm
$D$ and strange charm $D_S$ have been calculated. The open charmed meson masses in the eLSM (\ref{lag}) have been computed, which being in reasonably good
agreement with experimental data \cite{PDG}. We have evaluated the (OZI-dominant) decays of open
charmed mesons. The results are compatible with the results and the upper bounds listed by the PDG \cite{PDG}. Moreover, the decay
of the vector and axial-vector chiral partners $D^{\ast }(2010)$ and $%
D_{1}(2420)$ are well described. This fact shows that chiral symmetry is
still important for charmed mesons.\\

Further applications of the described approach are to calculate
the mixing of axial-vector and pseudovector charmed states and the
decay widths of hidden charmed mesons into light mesons, scalar
glueball, and pseudoscalar glueball. These works are currently in
progress.

\section*{Acknowledgments}

The author thanks F. Giacosa and D. H. Rischke for cooperation. Financial support from the Deutscher Akademischer Austausch Dienst (DAAD) is acknowledged.

\end{document}